\documentclass[12pt]{article}

\begin{document}
\begin{center}
{\bf Dirac-K\"ahler Equation}\footnote{Review}\\
\vspace{5mm}
 S.I. Kruglov \\
\vspace{5mm}
\textit{International Educational Centre, 2727 Steeles Ave. W, \# 202, \\
Toronto, Ontario, Canada M3J 3G9}
\end{center}

\begin{abstract}
Tensor and matrix formulations of Dirac-K\"ahler equation for massive and
massless fields are considered. The equation matrices obtained are simple
linear combinations of matrix elements in the 16-dimensional space. The
projection matrix-dyads defining all the 16 independent equation solutions
are found. A method of computing the traces of 16-dimensional
Petiau-Duffin-Kemmer matrix product is considered. It is shown that the
symmetry group of the Dirac-K\"ahler tensor fields is $SO(4,2)$. The
conservation currents corresponding this symmetry are constructed.
Supersymmetry of the Dirac-K\"ahler fields with tensor and spinor parameters
is analyzed. We show the possibility of constructing a gauge model of
interacting Dirac-K\"ahler fields where the gauge group is the noncompact
group under consideration.
\end{abstract}

\section{Introduction}

The important problems of particle physics are the confinement of quarks and
the chiral symmetry breaking (CSB) [1]. Both problems can not be solved
within perturbative quantum chromodynamics (QCD).

One of the promising methods in the infra-red limit of QCD is lattice QCD.
Lattice QCD takes into account both nonperturbative effects - CSB and the
confinement of quarks, and provides computational hadronic characteristics
with good accuracy. A natural framework of the lattice fermion formulation
and some version of Kogut-Suskind fermions [2, 3] are Dirac-K\"ahler
fermions [4-11]. The interest in this theory is due to the possibility of
applying the Dirac-K\"ahler equation for describing fermion fields with spin
$1/2$ on the lattice [4-11].

Recently much attention has been paid to the study of the Dirac-K\"ahler
field in the framework of differential forms [11-24]. The author [12]
considered an equation for inhomogeneous differential forms what is
equivalent to introducing a set of antisymmetric tensor fields of arbitrary
rank. It implies the simultaneous consideration of fields with different
spins.

K\"ahler [12] showed that the Dirac equation for particles with spin $1/2$
can be constructed from inhomogeneous differential forms. Now such fields
are called Dirac-K\"ahler fields. Using the language of differential forms,
Dirac-K\"ahler's equation in four dimensional space-time is given by
\begin{equation}
\left( d-\delta +m\right) \Phi =0  \label{1}
\end{equation}

where $d$ denotes the exterior derivative, $\delta =-\star d\star
$ turns $ n- $forms into $(n-1)-$form; $\star $ is the Hodge
operator [25] which connects a $n-$form to a $(4-n)-$form so, that
$\star ^2=1$, $d^2=\delta ^2=0 $. The Laplacian is given by
\[
\left( d-\delta \right) ^2=-\left( d\delta +\delta d\right)
=\partial _\mu \partial ^\mu
\]

So, the operator $\left( d-\delta \right) $ is the analog of the
Dirac operator $\gamma _\mu \partial ^\mu $. The inhomogeneous
differential form $ \Phi $ can be expanded as
\[
\Phi =\varphi (x)+\varphi _\mu (x)dx^\mu +\frac 1{2!}\varphi _{\mu \nu
}(x)dx^\mu \wedge dx^\nu +
\]
\begin{equation}
+\frac 1{3!}\varphi _{\mu \nu \rho }(x)dx^\mu \wedge dx^\nu \wedge dx^\rho
+\frac 1{4!}\varphi _{\mu \nu \rho \sigma }(x)dx^\mu \wedge dx^\nu \wedge
dx^\rho \wedge dx^\sigma  \label{2}
\end{equation}

where $\wedge $ is the exterior product. The form $\Phi $ includes
scalar $ \varphi (x)$, vector $\varphi _\mu (x)$, and
antisymmetric tensors $\varphi _{\mu \nu }(x)$, $\varphi _{\mu \nu
\rho }(x)$, $\varphi _{\mu \nu \rho \sigma }(x)$ fields. The
antisymmetric tensors of the third and forth ranks $ \varphi _{\mu
\nu \rho }(x)$, $\varphi _{\mu \nu \rho \sigma }(x)$ define a
pseudovector and pseudoscalar, respectively:
\begin{equation}
\widetilde{\varphi }_\mu (x)=\frac 1{3!}\epsilon _\mu ^{\nu \rho \sigma
}\varphi _{\nu \rho \sigma }(x),\hspace{0.5in}\widetilde{\varphi }(x)=\frac
1{4!}\epsilon ^{\mu \nu \rho \sigma }\varphi _{\mu \nu \rho \sigma }(x)
\label{3}
\end{equation}

In fact, the Dirac-K\"ahler equation (1) describes scalar, vector,
pseudoscalar and pseudovector fields. Some authors (see e. g. [6, 18-21])
showed that Eq. (1) is equivalent to four Dirac equations
\begin{equation}
\left( \gamma _\mu \partial _\mu +m\right) \psi
^{(b)}(x)=0,\hspace{0.5in} b=1,2,3,4 \label{4}
\end{equation}

The mapping between equations (1) and (4) makes it possible to describe
fermions with spin $1/2$ with the help of Eq. (1), i.e., boson fields ! As
we have already mentioned, this possibility is used in the lattice
formulation of QCD and for describing fermions with spin-$1/2$.

It should be noted that Ivanenko and Landau [25] considered (in 1928 ) an
equation for the set of antisymmetric tensor fields which is equivalent to
the Dirac-K\"ahler equation (1). Similar equations were discussed after
[26-32] long before the appearance of [12], and later rediscovered [33, 34].
The author [35, 36] found the internal symmetry group $SO(4,2)$ (or locally
isomorphic group $SU(2,2)$) of the Dirac-K\"ahler action and the
corresponding conserving currents. The transformations of the $SO(4,2)$
group mix the fields with different spins and do not commute with the
Lorentz transformations. Later others [37, 38, 18-21] also paid attention to
this symmetry. The transformations of the Lorentz and internal symmetry
groups discussed do not commute each other. So, parameters of the group are
tensors but not scalars as in (the more common) gauge theories. This kind of
symmetry is also different from supersymmetry where group parameters are not
scalars. The difference is that in our case the algebra of generators of the
symmetry is closed without adding the generators of the Poincar\'e group.
However the indefinite metric should be introduced here. The localization of
parameters of the internal symmetry group leads to the gauge fields and
field interactions.

In Section 2 we investigate the tensor and matrix formulations of
Dirac-K\"ahler equation for massive and massless fields. All
independent solutions of equations are found in the form of
matrix-dyads. It is shown in Section 3 that the internal symmetry
group of the Dirac-K\"ahler vector fields is $SO(4,2)$. For the
case of spinor fields we come to the $U(4)$ -group of symmetry.
The quantization of Dirac-K\"ahler's fields is carried out in
Section 4 by using an indefinite metric. It is shown in Section 5
that in field theory including Dirac-K\"ahler fields it is
possible to analyze supersymmetry groups with tensor and spinor
parameters without including coordinate transformations at the
same time. We show in Section 6 the possibility of constructing a
gauge model of interacting Dirac-K\"ahler fields where the gauge
group is the noncompact group $SO(4,2)$ under consideration.
Section 7. contains a conclusion. A method of computing the traces
of $16-$dimensional Petiau-Duffin-Kemmer matrix products is
considered in Appendix.

We use the system of units $\hbar =c=1$, $\alpha =e^2/4\pi
=1/137$, $e>0$, and Euclidean metrics, so that the squared
four-vector is $v_\mu ^2=\mathbf{v }^2+v_4^2=\mathbf{v}^2-v_0^2$
($\mathbf{v}^2=v_1^2+v_2^2+v_3^2$, $v_4=iv_0$).

\section{Tensor and matrix formulations of Dirac-K\"ahler
equation}

It is easy to show that Dirac-K\"ahler equation (1) with definitions (4) is
equivalent to the following tensor equations
\begin{equation}
\partial _\nu \varphi _{\mu \nu }+\partial _\mu \varphi +m\varphi _\mu =0
\hspace{0.5in}\partial _\nu \widetilde{\varphi }_{\mu \nu }+\partial _\mu
\widetilde{\varphi }+m\widetilde{\varphi }_\mu =0  \label{5}
\end{equation}
\begin{equation}
\partial _\mu \varphi _\mu +\varphi =0\hspace{0.3in}\partial _\mu \widetilde{
\varphi }_\mu +\widetilde{\varphi }=0  \label{6}
\end{equation}
\begin{equation}
\varphi _{\mu \nu }=\partial _\mu \varphi _\nu -\partial _\nu
\varphi _\mu -\varepsilon _{\mu \nu \alpha \beta }\partial _\alpha
\widetilde{\varphi } _\beta  \label{7}
\end{equation}

where
\begin{equation}
\widetilde{\varphi }_{\mu \nu }=\frac 12\varepsilon _{\mu \nu \alpha \beta
}\varphi _{\alpha \beta }  \label{8}
\end{equation}

is the dual tensor, $\varepsilon _{\mu \nu \alpha \beta }$ is an
antisymmetric tensor Levy-Civita; $\varepsilon _{1234}=-i$. It should be
noted that Eq. (7) is the most general representation for the antisymmetric
tensor of second rank in accordance with the Hodge theorem [38] (see also
[40-42]).

If the $\varphi $, $\widetilde{\varphi }$, $\varphi _\mu $,
$\widetilde{ \varphi }_\mu $ and $\varphi _{\mu \nu }$ are complex
values, Eqs. (5)-(7) describe the charged vector fields. These
equations are the tensor form of the Dirac-K\"ahler equation (1)
which was written in differential forms [12].

Now we show that in matrix form equations (5)-(7) are Dirac-like equations
with $16\times 16$ -dimensional Dirac matrices. The projection matrix-dyads
defining all the $16$ independent equation solutions will be constructed
[43].

The matrix form facilitates the investigation of the general group of
internal symmetry. To obtain the matrix form for both the massive and
massless cases, generalized equations are introduced:
\[
\partial _\mu \widetilde{\psi }_\mu +m_2\widetilde{\psi }_0=0
\]
\[
\partial _\nu \psi _{[\mu \nu ]}+\partial _\mu \psi _0+m_1\psi _\mu =0
\]
\[
\partial _\nu \psi _\mu -\partial _\mu \psi _\nu -e_{\mu \nu \alpha \beta
}\partial _\alpha \widetilde{\psi }_\beta +m_2\psi _{[\mu \nu ]}=0
\]
\begin{equation}
\partial _\mu \psi _\mu +m_2\psi _0=0  \label{9}
\end{equation}

With $m_1=m_2=m,$ $\psi _0=\varphi $, $\psi _\mu =\varphi _\mu $,
$\psi _{[\mu \nu ]}=\varphi _{\mu \nu },$ $\widetilde{\psi }_\mu
=i\widetilde{ \varphi }_\mu ,$ $\widetilde{\psi
}_0=i\widetilde{\varphi },$ $e_{\mu \nu \alpha \beta
}=i\varepsilon _{\mu \nu \alpha \beta }$ ($e_{1234}=1$), we arrive
at the Dirac-K\"ahler equations (5)-(8). In the case $m_1=0$, $
m_2\neq 0$ (where $m_2$ is the dimension parameter), Eqs. (9) are
the generalized Maxwell equations in the dual-symmetric form [44].
Let us introduce the $16-$component wave function

\begin{equation}
\Psi (x)=\left\{ \psi _A\right\} \hspace{0.5in}A=0,\mu ,[\mu \nu
],\widetilde{\mu },\widetilde{0} \label{10}
\end{equation}

It is convenient to introduce the matrix $\varepsilon ^{A,B}$ [45] with
dimension $16\times 16$; its elements consist of zeroes and only one element
is unity where row $A$ and column $B$ cross. Thus the multiplication and
matrix elements of these matrices are
\begin{equation}
\varepsilon ^{A,B}\varepsilon ^{C,D}=\varepsilon ^{A,D}\delta
_{BC} \hspace{0.5in}\left( \varepsilon ^{A,B}\right) _{CD}=\delta
_{AC}\delta _{BD} \label{11}
\end{equation}

where indexes $A,B,C,D=1,2,...,16$. Using the elements of the entire algebra
$\varepsilon ^{A,B},$ equations (9) take the form
\[
\biggl \{\partial _\nu \biggl [\varepsilon ^{\mu ,[\mu \nu
]}+\varepsilon ^{[\mu \nu ],\mu }+\varepsilon ^{\nu
,0}+\varepsilon ^{0,\nu }+\varepsilon ^{ \widetilde{\nu
},\widetilde{0}}+\varepsilon ^{\widetilde{0},\widetilde{\nu } }+
\]
\[
+\frac 12e_{\mu \nu \rho \omega }\left( \varepsilon
^{\widetilde{\mu },[\rho \omega ]}+\varepsilon ^{[\rho \omega
],\widetilde{\mu }}\right) \biggr ] _{AB}+
\]
\begin{equation}
+\left[ m_1\left( \varepsilon ^{\mu ,\mu }+\varepsilon
^{\widetilde{\mu }, \widetilde{\mu }}\right) +m_2\left(
\varepsilon ^{0,0}+\frac 12\varepsilon ^{[\mu \nu ],[\mu \nu
]}+\varepsilon ^{\widetilde{0},\widetilde{0}}\right) \right]
_{AB}\Psi _B(x)\biggr \}=0  \label{12}
\end{equation}

Let us introduce the projection matrices
\begin{equation}
\overline{P}=\varepsilon ^{\mu ,\mu }+\varepsilon ^{\widetilde{\mu
}, \widetilde{\mu }}\hspace{0.5in}P=\varepsilon ^{0,0}+\frac
12\varepsilon ^{[\mu \nu ],[\mu \nu ]}+\varepsilon
^{\widetilde{0},\widetilde{0}} \label{13}
\end{equation}

with the properties $P\overline{P}=\overline{P}P=0$,
$P+\overline{P}=I_{16}$ , and the $16\times 16-$matrix
\begin{equation}
\Gamma _\nu =\varepsilon ^{\mu ,[\mu \nu ]}+\varepsilon ^{[\mu \nu
],\mu }+\varepsilon ^{\nu ,0}+\varepsilon ^{o,\nu }+\varepsilon
^{\widetilde{\nu }, \widetilde{0}}+\varepsilon
^{\widetilde{0},\widetilde{\nu }}+\frac 12e_{\mu \nu \rho \omega
}\left( \varepsilon ^{\widetilde{\mu },[\rho \omega ]}+\varepsilon
^{[\rho \omega ],\widetilde{\mu }}\right)  \label{14}
\end{equation}

Then Eq. (12) takes the form of the relativistic wave equation
\begin{equation}
\left( \Gamma _\nu \partial _\nu +m_1\overline{P}+m_2P\right) \Psi (x)=0
\label{15}
\end{equation}

which includes both the massive and massless cases. The matrix $\Gamma _\nu $
can be represented in the form
\[
\Gamma _\nu =\beta _\nu ^{(+)}+\beta _\nu ^{(-)}\hspace{0.3in}\beta _\nu
^{(+)}=\beta _\nu ^{(1)}+\beta _\nu ^{(\widetilde{0})}\hspace{0.3in}\beta
_\nu ^{(-)}=\beta _\nu ^{(\widetilde{1})}+\beta _\nu ^{(0)}
\]
\[
\beta _\nu ^{(1)}=\varepsilon ^{\mu ,[\mu \nu ]}+\varepsilon ^{[\mu \nu
],\mu }\hspace{0.3in}\beta _\nu ^{(\widetilde{1})}=\frac 12e_{\mu \nu \rho
\omega }\left( \varepsilon ^{\widetilde{\mu },[\rho \omega ]}+\varepsilon
^{[\rho \omega ],\widetilde{\mu }}\right)
\]
\begin{equation}
\beta _\nu ^{(\widetilde{0})}=\varepsilon ^{\widetilde{\nu
},\widetilde{0} }+\varepsilon ^{\widetilde{0},\widetilde{\nu
}}\hspace{0.3in}\beta _\nu ^{(0)}=\varepsilon ^{\nu
,0}+\varepsilon ^{0,\nu }  \label{16}
\end{equation}

Matrices $\beta _\nu ^{(1)}$, $\beta _\nu ^{(\widetilde{1})}$ and
$\beta _\nu ^{(0)}$, $\beta _\nu ^{(\widetilde{0})}$ realize $10-$
and $5-$ dimensional irreducible representations of the
Petiau-Duffin-Kemmer [46-48] algebra
\begin{equation}
\beta _\mu ^{(1,0)}\beta _\nu ^{(1,0)}\beta _\alpha ^{(1,0)}+\beta _\alpha
^{(1,0)}\beta _\nu ^{(1,0)}\beta _\mu ^{(1,0)}=\delta _{\mu \nu }\beta
_\alpha ^{(1,0)}+\delta _{\alpha \nu }\beta _\mu ^{(1,0)}  \label{17}
\end{equation}
and $\beta _\nu ^{(+)}$, $\beta _\nu ^{(-)}$ are $16-$dimensional reducible
representations of the Petiau-Duffin-Kemmer algebra [28-31]. These matrices
obey the Petiau-Duffin-Kemmer algebra (17) and the matrix $\Gamma _\nu $ is
a $16\times 16-$Dirac matrix with the algebra:
\begin{equation}
\Gamma _\nu \Gamma _\mu +\Gamma _\mu \Gamma _\nu =2\delta _{\mu
\nu } \label{18}
\end{equation}

For the massive case when $m_1=m_2=m$, Eq. (15) becomes
\begin{equation}
\left( \Gamma _\nu \partial _\nu +m\right) \Psi (x)=0  \label{19}
\end{equation}

The $16-$component wave equation in the form of the first-order Eq. (19) was
also studied in [28-31, 33]. Now we find all independent solutions of Eq.
(19) in the form of matrix-dyads. In the momentum space Eq. (19) becomes
\begin{equation}
-i\widehat{p}\Psi _p=\varepsilon m\Psi _p  \label{20}
\end{equation}

where $\widehat{p}=\Gamma _\nu p_\nu $, and parameter $\varepsilon
=\pm 1$ corresponds to two values of the energy. From the property
of the Dirac matrices, Eq. (18), we find the minimal equation for
the operator $\widehat{p }$:
\begin{equation}
\left( i\widehat{p}+m\right) \left( i\widehat{p}-m\right) =0  \label{21}
\end{equation}

According to the general method [49, 50], the projection operator extracting
the states with definite energy (for particle or antiparticle) is given by
\begin{equation}
M_\varepsilon =\frac{m-i\varepsilon \widehat{p}}{2m}  \label{22}
\end{equation}

This operator has virtually the same form as in the Dirac theory of
particles with spin $1/2$. This is because the algebra of the matrices (18)
coincides with the algebra of the Dirac matrices $\gamma _\mu .$ However,
here we have the wave function $\Psi (x)$ which is transformed in the tensor
representation of the Lorentz group. It is also possible to use equation
(19) to describe spinor particles. In this case the wave function $\Psi (x)$
will be a spinor representation of the Lorentz group and equation (19) is
the direct sum of four Dirac equations (see (4)). This case is used for
fermions on the lattice [4-11].

Now we consider the bosonic case. To construct the spin operators we need
the generators of the Lorentz representation in the $16-$dimensional space
of the wave functions $\Psi (x)$. These generators are given in [28-31, 33]:
\begin{equation}
J_{\mu \nu }=\frac 14\left( \Gamma _\mu \Gamma _\nu -\Gamma _\nu \Gamma _\mu
+\overline{\Gamma }_\mu \overline{\Gamma }_\nu -\overline{\Gamma }_\nu
\overline{\Gamma }_\mu \right)  \label{23}
\end{equation}

where the matrices $\overline{\Gamma }_\nu $ also obey the Dirac algebra
(18) and have the form (see (16)):
\begin{equation}
\overline{\Gamma }_\nu =\beta _\nu ^{(+)}-\beta _\nu ^{(-)}  \label{24}
\end{equation}

It may be verified that the matrices $\Gamma _\mu $ and
$\overline{\Gamma } _\nu $ commute each other, i.e.
\begin{equation}
\left[ \Gamma _\mu ,\overline{\Gamma }_\nu \right] =0  \label{25}
\end{equation}

The spin projection operator here is given by
\begin{equation}
\sigma _p=-\frac i{2\mid \mathbf{p}\mid }\epsilon _{abc}p_aJ_{bc}=-\frac
i{4\mid \mathbf{p}\mid }\epsilon _{abc}p_a\left( \Gamma _b\Gamma _c+%
\overline{\Gamma }_b\overline{\Gamma }_c\right)  \label{26}
\end{equation}

which satisfies the following equation
\begin{equation}
\sigma _p\left( \sigma _p-1\right) \left( \sigma _p+1\right) =0  \label{27}
\end{equation}

In accordance with [49, 50] the corresponding projection operators are given
by
\begin{equation}
\widehat{S}_{(\pm 1)}=\frac 12\sigma _p\left( \sigma _p\pm
1\right) \hspace{0.5in}\widehat{S}_{(0)}=1-\sigma _p^2  \label{28}
\end{equation}

Operators $S_{(\pm 1)}$ correspond to the spin projections
$s_p=\pm 1$ and $ S_{(0)}$ to $s_p=0$. It is easy to verify that
the required commutation relations hold: $\widehat{S}_{(\pm
1)}^2=\widehat{S}_{(\pm 1)}$, $\widehat{S} _{(\pm
1)}\widehat{S}_{(0)}=0$, $\widehat{S}_{(0)}^2=\widehat{S}_{(0)}$.
The squared Pauli-Lubanski vector $\sigma ^2$ is given by
\begin{equation}
\sigma ^2=\left( \frac 1{2m}\varepsilon _{\mu \nu \alpha \beta }p_\nu
J_{\alpha \beta }\right) ^2=\frac 1{m^2}\left( J_{\mu \nu }^2p^2-J_{\mu
\sigma }J_{\nu \sigma }p_\mu p_\nu \right)  \label{29}
\end{equation}

It may be verified that this operator obeys the minimal equation
\begin{equation}
\sigma ^2\left( \sigma ^2-2\right) =0  \label{30}
\end{equation}

so that eigenvalues of the squared spin operator $\sigma ^2$ are $s(s+1)=0$
and $s(s+1)=2$. This confirms that the considered fields describe the
superposition of two spins $s=0$ and $s=1$. To separate these states we use
the projection operators
\begin{equation}
S_{(0)}^2=1-\frac{\sigma ^2}2\hspace{0.5in}S_{(1)}^2=\frac{\sigma ^2}2
\label{31}
\end{equation}

having the properties $S_{(0)}^2S_{(1)}^2=0$, $\left( S_{(0)}^2\right)
^2=S_{(0)}^2$, $\left( S_{(1)}^2\right) ^2=S_{(1)}^2$, $%
S_{(0)}^2+S_{(1)}^2=1 $, where $1\equiv I_{16}$ is the unit matrix
in $16-$ dimensional space. In accordance with the general
properties of the projection operators, the matrices $S_{(0)}^2$,
$S_{(1)}^2$ acting on the wave function extract pure states with
spin $0$ and $1$, respectively. Here there is a doubling of the
spin states of fields because we have scalar $ \psi _0$,
pseudoscalar $\widetilde{\psi }_0$, vector $\psi _\mu $, and
pseudovector $\widetilde{\psi }_\mu $, fields. To separate these
states it is necessary to introduce additional projection
operators. We use the following projection operator
\begin{equation}
\overline{M}_{\overline{\varepsilon
}}=\frac{m-i\overline{\varepsilon } \overline{p}}{2m}  \label{32}
\end{equation}

which has the same structure as Eq. (22) but with the matrix
$\overline{p}= \overline{\Gamma }_\nu p_\nu $ and an additional
quantum number $\overline{ \varepsilon }=\pm 1$. Following the
procedure [49, 50], 16 independent solutions in the form of
projection matrix-dyads are given by
\[
\Delta _{\varepsilon ,\pm 1,\overline{\varepsilon }}=\frac{\sigma
^2}2\cdot \frac{m-i\varepsilon \widehat{p}}{2m}\cdot
\frac{m-i\overline{\varepsilon } \overline{p}}{2m}\cdot \frac
12\sigma _p\left( \sigma _p\pm 1\right) =\Psi _{\varepsilon ,\pm
1,\overline{\varepsilon }}\cdot \overline{\Psi } _{\varepsilon
,\pm 1,\overline{\varepsilon }}
\]
\[
\Delta _{\varepsilon ,\overline{\varepsilon }}^{(1)}=\frac{\sigma
^2}2\cdot \frac{m-i\varepsilon \widehat{p}}{2m}\cdot
\frac{m-i\overline{\varepsilon } \overline{p}}{2m}\cdot \left(
1-\sigma _p^2\right) =\Psi _{\varepsilon , \overline{\varepsilon
}}\cdot \overline{\Psi }_{\varepsilon ,\overline{ \varepsilon }}
\]
\begin{equation}
\Delta _{\varepsilon ,\overline{\varepsilon }}^{(0)}=\left(
1-\frac{\sigma ^2 }2\right) \cdot \frac{m-i\varepsilon
\widehat{p}}{2m}\cdot \frac{m-i \overline{\varepsilon
}\overline{p}}{2m}\cdot \left( 1-\sigma _p^2\right) =\Psi
_{\varepsilon ,,\overline{\varepsilon }}^{(0)}\cdot \overline{\Psi
} _{\varepsilon ,\overline{\varepsilon }}^{(0)}  \label{33}
\end{equation}

where operators $\Delta _{\varepsilon ,\pm 1,\overline{\varepsilon
}}$, $ \Delta _{\varepsilon ,\overline{\varepsilon }}^{(1)}$
correspond to states with spin $1$ and spin projections $\pm 1$
and $0$, respectively and the projection operator $\Delta
_{\varepsilon ,\overline{\varepsilon }}^{(0)}$ extracts spin $0$.
The wave function $\Psi _p$ ($\Psi _{\varepsilon ,\pm 1,
\overline{\varepsilon }}$ or $\Psi _{\varepsilon
,\overline{\varepsilon }}$ or $\Psi _{\varepsilon
,,\overline{\varepsilon }}^{(0)}$) is the eigenvector of the
equations
\[
-i\widehat{p}\Psi _p=\varepsilon m\Psi
_p\hspace{0.5in}-i\overline{p}\Psi _p= \overline{\varepsilon
}m\Psi _p
\]
\begin{equation}
\sigma _p\Psi _p=s_p\Psi _p\hspace{0.5in}\sigma ^2\Psi _p=s(s+1)\Psi _p
\label{34}
\end{equation}

where the spin projections are $s_p=\pm 1,0$ and the spin is $s=1,0$. The
Hermitianizing matrix, $\eta $, in $16-$dimensional space is given by:
\begin{equation}
\eta =\Gamma _4\overline{\Gamma }_4  \label{35}
\end{equation}

This matrix obeys the equations
\[
\eta \Gamma _i=-\Gamma _i\eta \hspace{0.3in}(i=1,2,3)\hspace{0.3in}\eta
\Gamma _4=\Gamma _4\eta
\]
which guarantee the existence of a relativistically invariant bilinear form
\begin{equation}
\overline{\Psi }\Psi =\Psi ^{+}\eta \Psi  \label{36}
\end{equation}

where $\overline{\Psi }_p=\Psi ^{+}\Gamma _4\overline{\Gamma }_4$, and $\Psi
^{+}$ is the Hermitian-conjugate wave function.

In the spinor case, when Eq. (12) is the direct sum of four Dirac equations,
generators of the Lorentz group in $16-$dimensional space are
\begin{equation}
J_{\mu \nu }^{(1/2)}=\frac 14\left( \Gamma _\mu \Gamma _\nu -\Gamma _\nu
\Gamma _\mu \right)  \label{37}
\end{equation}

and the Hermitianizing matrix is $\eta _{^{1/2}}=\Gamma _4$. Using the
unitary transformation we can find the representation $\Gamma _\mu ^{\prime
}=I_4\otimes \gamma _\mu $, where $I_4$ is $4\times 4-$unit matrix, $\gamma
_\mu $ are the Dirac matrices and $\otimes $ means direct product. In this
basis the matrices $\overline{\Gamma }_\mu $ become $\overline{\Gamma }_\mu
^{\prime }=\gamma _\mu \otimes I_4$.

It is convenient also to use equations
\begin{equation}
\left( m-i\varepsilon \widehat{p}\right) \left(
m-i\overline{\varepsilon } \overline{p}\right) =2ip^{(\pm )}\left(
ip^{(\pm )}-\varepsilon m\right) \label{38}
\end{equation}
\begin{equation}
\sigma _p=\sigma _p^{(+)}=-\frac i{\mid \mathbf{p}\mid }\epsilon
_{abc}p_a\beta _b^{(+)}\beta _c^{(+)}=\sigma _p^{(-)}=-\frac
i{\mid \mathbf{p }\mid }\epsilon _{abc}p_a\beta _b^{(-)}\beta
_c^{(-)}  \label{39}
\end{equation}

where the sign ($+$) in Eq. (38) corresponds to the equality
$\varepsilon = \overline{\varepsilon }$, and sign ($-$) to
$\varepsilon =-\overline{ \varepsilon }$. With the help of Eqs.
(38), (39), the projection operators (33) are rewritten as
\[
\Delta _{\varepsilon ,\pm 1,\overline{\varepsilon }}=\frac{\sigma
^2}{8m^2} ip^{(\pm )}\left( ip^{(\pm )}-\varepsilon m\right)
\sigma _p^{(\pm )}\left( \sigma _p^{(\pm )}\pm 1\right) =\Psi
_{\varepsilon ,\pm 1}^{(\pm )}\cdot \overline{\Psi }_{\varepsilon
,\pm 1}^{(\pm )}
\]
\[
\Delta _{\varepsilon ,\overline{\varepsilon }}^{(1)}=\frac{\sigma
^2}{4m^2} ip^{(\pm )}\left( ip^{(\pm )}-\varepsilon m\right)
\left( 1-\sigma _p^{(\pm )2}\right) =\Psi _{\varepsilon
,\overline{\varepsilon }}^{(\pm )}\cdot \overline{\Psi
}_{\varepsilon ,\overline{\varepsilon }}^{(\pm )}
\]
\begin{equation}
\Delta _{\varepsilon ,\overline{\varepsilon }}^{(0)}=\frac
1{2m^2}\left( 1- \frac{\sigma ^2}2\right) ip^{(\pm )}\left(
ip^{(\pm )}-\varepsilon m\right) \left( 1-\sigma _p^{(\pm
)2}\right) =\Psi _{\varepsilon ,,\overline{ \varepsilon }}^{(\pm
)}\cdot \overline{\Psi }_{\varepsilon ,\overline{ \varepsilon
}}^{(\pm )}  \label{40}
\end{equation}

where $p^{(\pm )}=p_\mu \beta _\mu ^{(\pm )}.$ Projection matrix-dyads (40)
extract solutions $\Psi _p^{(\pm )}$ which are the solutions of the
equations
\begin{equation}
-ip^{(+)}\Psi _p^{(+)}=\frac 12\left( \varepsilon
+\overline{\varepsilon } \right) m\Psi _p^{(+)}  \label{41}
\end{equation}
\begin{equation}
-ip^{(-)}\Psi _p^{(-)}=\frac 12\left( \varepsilon
-\overline{\varepsilon } \right) m\Psi _p^{(-)}  \label{42}
\end{equation}
\begin{equation}
\sigma _p^{(\pm )}\Psi _p^{(\pm )}=s_p\Psi _p^{(\pm )}\hspace{0.5in}\sigma
^2\Psi _p^{(\pm )}=s(s+1)\Psi _p^{(\pm )}  \label{43}
\end{equation}

With $\varepsilon =\overline{\varepsilon }$, Eq. (41) describes
the superposition of vector and pseudoscalar fields, and Eq. (42)
with $ \varepsilon =-\overline{\varepsilon }$ describes the
superposition of pseudovector and scalar fields (see [28-31]).

Let us investigate the case of massless fields; Eq. (15) with $m_1=0$
becomes
\begin{equation}
\left( \Gamma _\nu \partial _\nu +m_2P\right) \Psi (x)=0  \label{44}
\end{equation}

In the momentum space, the field function $\Psi _k$ is the solution to the
equation
\begin{equation}
B\Psi _k=0\hspace{0.5in}B=i\widehat{k}+m_2P  \label{45}
\end{equation}

where $\widehat{k}=\Gamma _\mu k_\mu $, $k_\mu ^2=0$ and the matrix $B$
obeys the minimal equation
\begin{equation}
B\left( B-m_2\right) =0  \label{46}
\end{equation}
The projection operator which extracts the solution to Eq. (45) is
\begin{equation}
\alpha =\frac{m_2-B}{m_2}  \label{47}
\end{equation}

with the equality $\alpha ^2=\alpha $ required by a projection operator. The
corresponding spin operators are given by
\begin{equation}
\sigma _k=-\frac i{k_0}\epsilon _{abc}k_a\beta _b^{(\pm )}\beta _c^{(\pm )}
\label{48}
\end{equation}

As a result, the projection matrix-dyads, corresponding to the generalized
Maxwell field after extracting spin $0$ and spin projections $\pm 1$, take
the form
\[
\Pi _{(0)}=\frac 1{m_2}\left( 1-\frac{\sigma ^2}2\right) \left( m_2-B\right)
=\Psi ^{_{(0)}}\cdot \overline{\Psi }^{_{(0)}}
\]
\begin{equation}
\Pi _{(\pm 1)}=\frac 1{2m_2}\sigma _k\left( \sigma _k\pm 1\right) \left(
m_2-B\right) =\Psi ^{_{(\pm )}}\cdot \overline{\Psi }^{_{(\pm )}}  \label{49}
\end{equation}

Let us consider the case of spinor particles when the wave
function $\Psi (x) $ realizes the spinor representation of the
Lorentz group with the generators (37) and Hermitianizing matrix
$\eta _{^{1/2}}=\Gamma _4$. In this case the wave function $\Psi
(x)$ represents the direct sum of four bispinors and the variables
$\psi _0$, $\psi _\mu $, $\psi _{[\mu \nu ]},$ $ \widetilde{\psi
}_\mu ,$ $\widetilde{\psi }_0$, which comprise $\Psi (x)$ (10),
are connected with components of spinors. Under the Lorentz
transformations with generators (37) these variables do not
transform as tensors. Thus the equations for the eigenvalues and
the spin operator are:
\[
-i\widehat{p}\Psi _p^{(1/2)}=\varepsilon m\Psi
_p^{(1/2)}\hspace{0.5in} \sigma _p^{(1/2)}\Psi _p^{(1/2)}=s_p\Psi
_p^{(1/2)}
\]
\begin{equation}
\sigma _p^{(1/2)}=-\frac i{4\mid \mathbf{p}\mid }\epsilon _{abc}p_a\Gamma
_b\Gamma _c  \label{50}
\end{equation}

where $s_p=\pm 1/2$. As there is a degeneracy of states due to the $16$ -
dimensionality we should use additional equations with the corresponding
quantum numbers. Taking into account Eq. (25) we can use the following
additional equations to separate states of spinor fields:
\[
-i\overline{p}\Psi _p^{(1/2)}=\overline{\varepsilon }m\Psi
_p^{(1/2)} \hspace{0.5in}\overline{\sigma }_p^{(1/2)}\Psi
_p^{(1/2)}=\overline{s}_p\Psi _p^{(1/2)}
\]
\begin{equation}
\overline{\sigma }_p^{(1/2)}=-\frac i{4\mid \mathbf{p}\mid }\epsilon
_{abc}p_a\overline{\Gamma }_b\overline{\Gamma }_c  \label{51}
\end{equation}

with $\overline{s}_p=\pm 1/2$. We can treat the additional quantum
number $ \overline{s}_p$ as the ``internal spin'' because matrices
(24) obey the Dirac algebra
\begin{equation}
\overline{\Gamma }_\nu \overline{\Gamma }_\mu +\overline{\Gamma
}_\mu \overline{\Gamma }_\nu =2\delta _{\mu \nu }  \label{52}
\end{equation}
Thus it is easy to find all independent solutions of equations (50), (51) in
the form of matrix-dyads:
\[
\Delta _{\varepsilon ,\overline{\varepsilon
},s_p,\overline{s}_p}=\frac{ m-i\varepsilon \widehat{p}}{2m}\cdot
\frac{m-i\overline{\varepsilon } \overline{p}}{2m}\cdot \left(
\frac 12+2s_p\sigma _p^{(1/2)}\right) \left( \frac
12+2\overline{s}_p\overline{\sigma }_p^{(1/2)}\right)
\]
\begin{equation}
=\Psi _{\varepsilon ,,\overline{\varepsilon },s_p,\overline{s}_p}\cdot
\overline{\Psi }_{\varepsilon ,\overline{\varepsilon },s_p,\overline{s}_p}
\label{53}
\end{equation}

where $\overline{\Psi }_{\varepsilon ,\overline{\varepsilon
},s_p,\overline{s }_p}=\Psi _{\varepsilon ,,\overline{\varepsilon
},s_p,\overline{s} _p}^{+}\Gamma _4$, $\varepsilon =\pm 1$,
$s_p=\pm 1/2$, and we introduce two additional internal quantum
numbers $\overline{\varepsilon }=\pm 1$, $ \overline{s}_p=\pm
1/2$. In [33] the author also used a similar construction for the
solutions of the field equations but without the dyad
representation (53). The dyad representation is essential as all
quantum electrodynamics calculations can only be done using
matrix-dyads [49, 50].

\section{Symmetry group $O(4.2)$ for charged vector fields}

The group $SO(4,2)$ is the symmetry group of the Dirac-K\"ahler vector
fields. This will be shown using the Dirac matrix algebra and the minimality
of the electromagnetic interaction [36].

The interaction with electromagnetic field is introduced by the
substitution $\partial _\mu \rightarrow D_\mu ^{(-)}=\partial _\mu
-ieA_\mu $, where $ A_\mu $ is the vector -potential of the
electromagnetic field. Consider the Lagrangian
\begin{equation}
\mathcal{L}=-\frac 12\left[ \overline{\Psi }(x)\left( \Gamma _\mu
\overrightarrow{D_\mu ^{(-)}}+m\right) \Psi (x)-\overline{\Psi
}(x)\left( \Gamma _\mu \overleftarrow{D_\mu ^{(+)}}-m\right) \Psi
(x)\right] -\frac 14 \mathcal{F}_{\mu \nu }  \label{54}
\end{equation}

where $D_\mu ^{(+)}=\partial _\mu +ieA_\mu $, $\overline{\Psi }=\Psi
^{+}\Gamma _4\overline{\Gamma }_4$, and the arrows above the $D_\mu ^{(\pm
)} $ show the direction in which these operators act; $\mathcal{F}_{\mu \nu
}=\partial _\mu A_\nu -\partial _\nu A_\mu $ is the strength tensor of the
electromagnetic field. From the variation of the Lagrangian (54) on wave
functions $\Psi $ and $\overline{\Psi },$ we find equations for
Dirac-K\"ahler vector fields in the external electromagnetic fields
\begin{equation}
\left( \Gamma _\mu D_\mu ^{(-)}+m\right) \Psi
(x)=0\hspace{0.5in}\overline{ \Psi }(x)\left( \Gamma _\mu
\overleftarrow{D_\mu ^{(+)}}-m\right) =0 \label{55}
\end{equation}
From the variation of Eq. (55) on the vector-potential $A_\mu $ we get
Maxwell equations, in which the source is the electromagnetic current $J_\mu
^{el}=ie\overline{\Psi }\Gamma _\mu \Psi $.

Let us consider the set of 16 linear independent matrices:
\[
I=I_{16}\hspace{0.3in}I_\mu =\overline{\Gamma }_\mu \hspace{0.3in}
I_{\mu \nu }=\frac 14\overline{\Gamma }_{_{[\mu }}\overline{\Gamma
} _{\nu ]}
\]
\begin{equation}
\widetilde{I}=\overline{\Gamma }_5=\overline{\Gamma
}_1\overline{\Gamma }_2 \overline{\Gamma }_3\overline{\Gamma
}_4\hspace{0.3in}\widetilde{I} _\mu =\overline{\Gamma
}_5\overline{\Gamma }_\mu  \label{56}
\end{equation}

They commute with the operators $\left( \Gamma _\mu D_\mu ^{(-)}+m\right) $,
$\left( \Gamma _\mu \overleftarrow{D_\mu ^{(+)}}-m\right) $ of Eqs. (55) and
generate the algebra of the symmetry of Eqs. (55). The algebra of the
generators (56) is isomorphic to the Clifford algebra with the commutation
relations:
\[
\left[ I_{\alpha \beta },I_{\mu \nu }\right] =\delta _{\beta \mu }I_{\alpha
\nu }+\delta _{\alpha \nu }I_{\beta \mu }-\delta _{\beta \nu }I_{\alpha \mu
}-\delta _{\alpha \mu }I_{\beta \nu }
\]
\[
\left[ I_\mu ,I_{\alpha \beta }\right] =\delta _{\mu \alpha }I_\beta -\delta
_{\mu \beta }I_\alpha
\]
\[
\left[ \widetilde{I}_\mu ,I_{\alpha \beta }\right] =\delta _{\mu
\alpha } \widetilde{I}_\beta -\delta _{\mu \beta
}\widetilde{I}_\alpha
\]
\[
\left[ \widetilde{I}_\mu ,I_\nu \right] =2\widetilde{I}\delta
_{\mu \nu } \hspace{0.3in}\left[ I_\mu ,I_\nu \right] =4I_{\mu \nu
}\hspace{0.3in}\left[ \widetilde{I}_\mu ,\widetilde{I}_\nu \right]
=-4I_{\mu \nu }
\]
\begin{equation}
\left[ I_\mu ,\widetilde{I}\right] =-2\widetilde{I}_\mu \hspace{0.3in}\left[
\widetilde{I}_\mu ,\widetilde{I}\right] =-2I_\mu \hspace{0.3in}\left[
I_{\alpha \beta },\widetilde{I}\right] =0  \label{57}
\end{equation}

Let us introduce the anti-Hermitian generators [51]:
\[
I_0=iI_{16}\hspace{0.3in}I_{56}=-I_{65}=\frac i2\widetilde{I}
\]
\begin{equation}
I_{6\mu }=-I_{\mu 6}=\frac 12\widetilde{I}_\mu \hspace{0.3in}I_{5\mu
}=-I_{\mu 5}=\frac i2I_\mu  \label{58}
\end{equation}

With the help of Eqs. (58), the commutation relations (57) take the form
\[
\left[ I_{AB},I_{CD}\right] =\delta _{BC}I_{AD}+\delta _{AD}I_{BC}-\delta
_{AC}I_{BD}-\delta _{BD}I_{AC}
\]
\begin{equation}
\left[ I_{AB},I_0\right] =0\hspace{0.3in}A,B,C,D,=1, 2,...,6
\label{59}
\end{equation}

The algebra (59) corresponds with the direct product of the group
of $6-$ dimensional rotation $SO(6,6)$ and the unitary group
$U(1)$ (for real group parameters). This group is isomorphic to
the $U(4)$ group. The transformations of the corresponding group
are given by
\[
\Psi ^{\prime }(x)=U\Psi (x)
\]
\begin{equation}
U=\exp \left( I\alpha +I_\mu \beta _\mu +I_{\mu \nu }\omega _{\mu
\nu }+ \widetilde{I}_\mu \delta _\mu +\widetilde{I}\xi \right)
\label{60}
\end{equation}

where $\alpha ,$ $\beta _\mu ,$ $\omega _{\mu \nu },$ $\delta _\mu
,$ $\xi $ are the group parameters; if these parameters are
complex, we have the $ GL(4,c)$ group. For the neutral
Dirac-K\"ahler fields, the transformations (60) should leave real
components as real components with the conditions: $ \alpha
^{*}=\alpha $, $\beta _m^{*}=\beta _m$, $\beta _4^{*}=-\beta _4$,
$ \omega _{mn}^{*}=\omega _{mn}$, $\omega _{m4}^{*}=-\omega
_{m4}$, $\delta _m^{*}=-\delta _m,$ $\delta _4^{*}=\delta _4,$
$\xi ^{*}=-\xi $ corresponding to the $SO(3,3)\otimes GL(1,R)$
group. Such a contraction of the $GL(4,c)$ group is a consequence
of charged fields being described by complex fields having more
degrees of freedom.

The requirement that the Lagrangian (54) is invariant under the
transformations (60) leads to the constraints: $\alpha
^{*}=-\alpha $, $ \beta _m^{*}=\beta _m$, $\beta _4^{*}=-\beta
_4$, $\omega _{mn}^{*}=\omega _{mn}$, $\omega _{m4}^{*}=-\omega
_{m4}$, $\delta _m^{*}=\delta _m,$ $\delta _4^{*}=-\delta _4$,
$\xi ^{*}=\xi $ which corresponds to contraction of the $
SO(4,2)\otimes U(1)$ group with 16 parameters [36]. This occurs
only for charged Dirac-K\"ahler fields. The subgroup $U(1)$ is the
known group of gauge transformations: $\Psi ^{\prime }(x)=\exp
\left( I\alpha \right) \Psi (x)$ ($\alpha ^{*}=-\alpha $) which
gives the conservation law of four-current. For neutral fields,
the group leaving the Lagrangian invariant under the
transformation is $SO(3,2)$ with generators $\widetilde{I}_\mu $,
$ I_{\mu \nu }$ and corresponding parameters $\omega
_{mn}^{*}=\omega _{mn}$, $ \omega _{m4}^{*}=-\omega _{m4}$,
$\delta _m^{*}=-\delta _m,$ $\delta _4^{*}=\delta _4$. The
generators $I_{\mu \nu }$ (see (57)) with parameters $ \omega
_{mn}^{*}=\omega _{mn}$, $\omega _{m4}^{*}=-\omega _{m4}$
correspond to the subgroup $SO(3,1)$. The one-parameter subgroup
of the Larmor transformations with the generator $\widetilde{I}$
was mentioned in [28-31]. Only generators of Larmor and gauge
transformations commute with the Lorentz group generators (23).
This means that in the general case, the transformations of the
group with internal symmetry $SO(4,2)$ do not commute with the
Lorentz transformations. The transformations of the Lorentz group
realize the operation of internal automorphism with respect to the
elements of the group considered. As a consequence, the parameters
of this group are tensors. This is the main difference between the
group being considered and the usual groups of internal symmetry
where parameters are scalars.

As the Lagrangian (54) is invariant under the $SO(4,2)\otimes U(1)$ group we
find in accordance with Noether's theorem that the variation of the action
is
\begin{equation}
\delta S=\int d^4x\partial _\mu \left( \overline{\Psi }(x)\Gamma _\mu \delta
\Psi (x)-\delta \overline{\Psi }(x)\Gamma _\mu \Psi (x)\right) =0  \label{61}
\end{equation}

As parameters of transformations (60) are independent we find from (61) the
conservation tensors:
\[
J_\mu =i\overline{\Psi }(x)\Gamma _\mu \Psi (x)\hspace{0.3in}K_\mu
= \overline{\Psi }(x)\Gamma _\mu \overline{\Gamma }_5\Psi
(x)\hspace{0.3in} R_{\mu \alpha }=\overline{\Psi }(x)\Gamma _\mu
\overline{\Gamma }_5\overline{ \Gamma }_\alpha \Psi (x)
\]
\begin{equation}
C_{\mu \alpha }=\overline{\Psi }(x)\Gamma _\mu \overline{\Gamma
}_\alpha \Psi (x)\hspace{0.3in}\Theta _{\mu [\alpha \beta
]}=\overline{\Psi } (x)\Gamma _\mu \overline{\Gamma }_{[\alpha
}\overline{\Gamma }_{\beta ]}\Psi (x)  \label{62}
\end{equation}

These conservation currents were also constructed in [28-31] without
consideration of the corresponding internal symmetry. Conservation of the
current (62) follows from the symmetry-group $SO(4,2)\otimes U(1)$ of the
Lagrangian (54). For the neutral Dirac-K\"ahler fields, there is a
conservation of tensors $C_{\mu \alpha }$ and $\Theta _{\mu [\alpha \beta ]}$
corresponding to the symmetry-subgroup $SO(3,2).$ Using the matrices $\Gamma
_\mu $, $\overline{\Gamma }_\mu $ (16), (24) and wave functions (10) $\Psi
(x)$, $\overline{\Psi }(x)=\Psi ^{+}(x)\Gamma _4\overline{\Gamma }_4$, it is
easy to verify that in this case (of neutral fields), the currents $J_\mu $,
$K_\mu $ and $R_{\mu \alpha }$ are identically zero.

For the spinor case with the generators (37), the Lagrangian (54)
with the conjugate function $\overline{\Psi }(x)=\Psi
^{+}(x)\Gamma _4$ is invariant under the $U(4)$ - group
transformation (60) with the parameter constraints: $\alpha
^{*}=-\alpha $, $\beta _\mu ^{*}=-\beta _\mu $, $\omega _{\mu \nu
}^{*}=\omega _{\mu \nu }$, $\delta _\mu ^{*}=\delta _\mu ,$ $\xi
^{*}=-\xi $ . In this case the transformation (60) commutes with
the Lorentz transformations (see (37). The existence of the
additional quantum numbers $ \overline{s}_p$ and
$\overline{\varepsilon }$ is connected here with the presence of
the group $SU(4)$. The subgroup $U(1)$ with the parameter $ \alpha
$ is the well known group of gauge transformations giving the
conservation of the electric current $J_\mu =i\overline{\Psi
}(x)\Gamma _\mu \Psi (x)$.

\section{Quantization of fields}

The quantization of Dirac-K\"ahler's fields will be carried out by using an
indefinite metric. It will be shown that the renormalization procedure is
carried out in the same manner as in quantum electrodynamics [52, 53].

The Lagrangian of charged Dirac-K\"ahler fields (54) to within
four-divergences can be written (when electromagnetic fields are absent) as
\begin{equation}
\mathcal{L}=-\overline{\Psi }(x)\left( \Gamma _\mu \partial _\mu +m\right)
\Psi (x)  \label{63}
\end{equation}

where $\overline{\Psi }(x)=\Psi (x)^{+}\Gamma _4\overline{\Gamma
}_4$ corresponds to the tensor representation of the Lorentz
group, where the 16-component wave function $\Psi $ describes
scalar, pseudoscalar, vector and pseudovector fields. In the case
of spinor representation of the Lorentz group, $\Psi $ is the
direct sum of four Dirac bispinors, $\overline{\Psi } (x)=\Psi
(x)^{+}\Gamma _4$, and the quantizing procedure is similar to the
Dirac theory.

Now we will consider the case of the boson fields. Using the canonical
quantization, one arrives at the commutators
\begin{equation}
\left[ \Psi _M(x),\overline{\Psi }_N(x^{\prime })\right] _{t=t^{\prime
}}=\left( \Gamma _4\right) _{MN}\delta (\mathbf{x}-\mathbf{x}^{\prime })
\label{64}
\end{equation}

where $M$, $N=1$,$2$,...,$16$. It follows from (4.2) that it is necessary to
introduce the indefinite metric, as for finite-dimensional equations (with
the diagonal matrix $\Gamma _4$) which describe fields with integer spins;
only fields obeying the Petiau-Duffin-Kemmer equation have positive energy
[54]. With the help of Eqs. (10), (14) we get the following commutation
relations for the tensor fields:
\[
\left[ \varphi (x),\varphi _0^{*}(x^{\prime })\right]
_{t=t^{\prime }}=i\delta (\mathbf{x}-\mathbf{x}^{\prime
})\hspace{0.3in}\left[ \widetilde{ \varphi }(x),\widetilde{\varphi
}_0^{*}(x^{\prime })\right] _{t=t^{\prime }}=-i\delta
(\mathbf{x}-\mathbf{x}^{\prime })
\]
\[
\left[ \varphi _k(x),\varphi _{n4}^{*}(x^{\prime })\right] _{t=t^{\prime
}}=\delta _{kn}\delta (\mathbf{x}-\mathbf{x}^{\prime })
\]
\begin{equation}
\left[ \widetilde{\varphi }_k(x),\varphi _{mn}^{*}(x^{\prime })\right]
_{t=t^{\prime }}=i\epsilon _{kmn}\delta (\mathbf{x}-\mathbf{x}^{\prime })
\label{65}
\end{equation}

plus complex conjugated relations. In the momentum space, the equation of
motion of fields with spins $0$ and $1$, takes the form
\begin{equation}
\left( m\pm i\widehat{p}\right) \Psi ^{\pm }(\mathbf{p})=0  \label{66}
\end{equation}

where $\widehat{p}=p_\mu \Gamma _\mu $, $\Psi ^{\pm }(\mathbf{p})$
are positive ($\Psi ^{+}(\mathbf{p})$) and negative ($\Psi
^{-}(\mathbf{p})$) frequency parts of the wave function
corresponding positive $p_0>0$ and negative $p_0<0$ energies of
particles, respectively. For each value of the energy, there are
eight solutions with definite spin, spin projection and addition
quantum number. Wave functions $\Psi ^{\pm }(\mathbf{p})$, $
\overline{\Psi }^{\pm }(\mathbf{p})$ can be expanded in spin
states as follows
\begin{equation}
\Psi _M^{\pm }(\mathbf{p})=a_s^{\mp }(\mathbf{p})v_M^{s,\mp
}(\mathbf{p}) \hspace{0.3in}\overline{\Psi }_M^{\pm
}(\mathbf{p})=a_s^{*\mp }(\mathbf{p} )v_N^{*s,\mp
}(\mathbf{p})\left( \Gamma _4\overline{\Gamma }_4\right) _{NM}
\label{67}
\end{equation}

where index $s=0,m,\widetilde{n},\widetilde{0}$ ($m$, $n=1,2,3$),
and operator $a_s^{*+}(\mathbf{p})$ is the creation operator of a
particle in the scalar state ($s=0$), pseudoscalar state
($s=\widetilde{0}$), vector state ($s=m$), pseudovector state
($s=\widetilde{n}$), and $a_s^{-}(\mathbf{p })$ is the
annihilation operator of a particle. The normalization conditions
for solutions (67) are different from the Dirac bispinors case,
and are given by
\begin{equation}
v_N^{*s,\pm }(\mathbf{p})\left( \overline{\Gamma }_4\right) _{NM}v_M^{r,\mp
}(\mathbf{p})=\pm \varepsilon _s\delta _{sr}  \label{68}
\end{equation}
\begin{equation}
\overline{v}^{s,\pm }(\mathbf{p})v^{r,\mp
}(\mathbf{p})=\varepsilon _s\delta _{sr}\frac
m{p_0}\hspace{0.3in}\left( \overline{v}^{s,\pm }(\mathbf{p}
)=v^{*s,\pm }(\mathbf{p})\Gamma _4\overline{\Gamma }_4\right)
\label{69}
\end{equation}
\begin{equation}
v^{*s,\pm }(\mathbf{p})\overline{\Gamma }_4v^{r,\pm
}(-\mathbf{p})=0 \hspace{0.3in}\left( v^{s,\pm
}(\mathbf{p})\right) ^{*}=v^{*s,\mp }(\mathbf{p })  \label{70}
\end{equation}

and $\varepsilon _s=1$ at $s=\widetilde{0},$ $m,$ and $\varepsilon
_s=-1$ at $s=0,$ $\widetilde{n}.$ We use here the normalization on
the charge and in the right hand sides of Eqs. (68), (69) there is
no summation on indexes $s$ . The summation formula on indexes $s$
corresponding to the normalization conditions, has the form
\begin{equation}
\sum_s\varepsilon _sv_M^{s,\pm }(\mathbf{p})\overline{v}_N^{s,\mp
}(\mathbf{p })=\left( \frac{m\pm i\widehat{p}}{2p_0}\right) _{MN}
\label{71}
\end{equation}

If we take the trace of the matrix (71) and compare it with the expression
(69) summed over all states $s$, we will get the equality. Multiplying Eq.
(71) into the matrix $\Gamma _4$, and then calculating the trace of both
sides of the matrix equality, we arrive at Eq. (68).

The appearance of the coefficient $\varepsilon _s=\pm 1$ in the right hand
side of Eq. (69) reflects the fact that the energy of vector and
pseudoscalar states is positive, and the energy of pseudovector and scalar
states is negative. We can also come to this conclusion using the expression
for the energy-momentum tensor
\begin{equation}
T_{\mu \nu }=-\overline{\Psi }(x)\Gamma _\mu \partial _\nu \Psi (x)
\label{72}
\end{equation}

which follows from the Lagrangian (63). Taking into account the expansions
for wave functions:
\[
\Psi (x)=(2\pi )^{-3/2}\int \left[ \Psi
^{+}(\mathbf{p})e^{ipx}+\Psi ^{-}( \mathbf{p})e^{-ipx}\right] d^3p
\]
\begin{equation}
\overline{\Psi }(x)=(2\pi )^{-3/2}\int \left[ \overline{\Psi
}^{+}(\mathbf{p} )e^{ipx}+\overline{\Psi
}^{-}(\mathbf{p})e^{-ipx}\right] d^3p  \label{73}
\end{equation}

found from Eq. (72), the energy-momentum-vector
\begin{equation}
P_\mu =-i\int \overline{\Psi }(x)\Gamma _4\partial _\mu \Psi (x)d^3x
\label{74}
\end{equation}

and Eqs. (67)-(70), we arrive at the following expression
\begin{equation}
P_\mu =\int p_\mu \sum_s\varepsilon _s\left(
a_s^{*+}(\mathbf{p})a_s^{-}(
\mathbf{p})+a_s^{*-}(\mathbf{p})a_s^{+}(\mathbf{p})\right) d^3p
\label{75}
\end{equation}

To have the positive energy in accordance with the Eq. (75), we should use
the following commutation relations for creation and annihilation operators:
\begin{equation}
\left[ a_s^{*-}(\mathbf{p}),a_r^{+}(\mathbf{p})\right] =\left[
a_s^{-}( \mathbf{p}),a_r^{*+}(\mathbf{p})\right] =\varepsilon
_s\delta _{sr}\delta ( \mathbf{p}-\mathbf{p}^{\prime })
\label{76}
\end{equation}

With the help of Eqs. (67), (71) and (76) we find
\[
\left[ \Psi _M^{-}(x),\overline{\Psi }_N^{+}(y)\right] =-(2\pi )^{-3}\int
\frac{\left( m+i\widehat{p}\right) _{MN}}{2p_0}e^{-ip(x-y)}d^3p
\]
\begin{equation}
=\left( \Gamma _\mu \frac \partial {\partial x_\mu }-m\right) _{MN}\Delta
_{-}(x-y)  \label{77}
\end{equation}
\[
\left[ \Psi _M^{+}(x),\overline{\Psi }_N^{-}(y)\right] =(2\pi )^{-3}\int
\frac{\left( m-i\widehat{p}\right) _{MN}}{2p_0}e^{ip(x-y)}d^3p
\]
\begin{equation}
=-\left( \Gamma _\mu \frac \partial {\partial x_\mu }-m\right) _{MN}\Delta
_{+}(x-y)  \label{78}
\end{equation}

It follows from Eqs. (77), (78) (by taking into account the definition of
the Pauli-Jordan function [55, 56] $\Delta _0=i\left( \Delta _{+}(x)-\Delta
_{-}(x)\right) $) that
\[
\left[ \Psi (x),\overline{\Psi }(y)\right] =S(x-y)\hspace{0.3in}
S(x-y)=S^{+}(x-y)+S^{-}(x-y)
\]
\begin{equation}
S^{\pm }(x-y)=\mp \left( \Gamma _\mu \frac \partial {\partial x_\mu
}-m\right) \Delta _{\pm }(x-y)  \label{79}
\end{equation}

where the function $S(x-y)$ satisfies the following equations
\begin{equation}
\left( \Gamma _\mu \frac \partial {\partial x_\mu }+m\right) S(x-y)=i\left(
\frac{\partial ^2}{\partial x_\mu ^2}-m^2\right) \Delta _0(x-y)=0  \label{80}
\end{equation}

From Eqs. (79), at $t=t^{\prime }$, and using Eqs. (77), (78) we arrive at
the commutator (64). With the help of the relationships (79), we can find
the chronological pairing of the operators:
\[
\langle T\Psi _M(x)\overline{\Psi }_N(y)\rangle _0=S_{MN}^c(x-y)
\]
\[
=\Theta (x_0-y_0)S_{MN}^{+}(x-y)-\Theta (y_0-x_0)S_{MN}^{-}(x-y)
\]
\begin{equation}
=\frac i{(2\pi )^4}\int \frac{\left( i\widehat{p}-m\right) _{MN}}{
p^2+m^2-i\varepsilon }e^{ip(x-y)}d^4p  \label{81}
\end{equation}

which has formally the same form as in quantum electrodynamics
(QED). Here $ \Theta (x)$ is the well known theta-function [55]
and $x_0$ is the time.

It is seen from Eq. (81), that the Feynman rules for particles with
multispin $0,1$ interacting with the electromagnetic field, eventually are
the same as in QED. We should not, however, here use the QED factor $\eta
=(-1)^l$, where $l$ is the number of loops in the diagram due to different
statistics [55]. The difference is in the number of spin states of the
charged particle, and in the dimension of matrices $\Gamma _\mu $. As the
propagator (81) formally coincides with the electron propagator of QED, so
all divergences can be cancelled by the standard procedure, i. e., we have
here a renormalizable theory. All matrix elements of quantum processes
describing the interaction of particles with multispin $0,1$ coincide
eventually with the corresponding elements in QED. The difference is in the
density matrix $\Psi \cdot \overline{\Psi }$ which we found in Sec. 2.

The commutation relations (76) with sign ($-$) in the right hand
side (at $ \varepsilon _s=-1$) require the introduction of the
indefinite metric. The space of states is divided into two
substates: $H_p$ and $H_n$ with positive ($H_p$) and negative
($H_n$) square norm. The vector and pseudoscalar states correspond
to a positive square norm, and pseudovector and scalar states $-$
to a negative square norm. The total space is the direct sum of
the two subspaces $H_p$ and $H_n$.

\section{Supersymmetry of Dirac-K\"ahler's fields}

For the Dirac-K\"ahler fields, it will be shown that in field theory it is
possible to analyze transformation groups with tensor and spinor parameters
without including coordinate transformations at the same time [57].

A graduated Lie algebra must be absolutely related to transformations
involving space-time coordinates [58], but it seems perfectly obvious that
this is always the case if we are dealing with transformations whose
generators are of neither a tensor nor spinor nature (see [59, 60], for
example).

A theory of Dirac-K\"ahler's fields, however, raises the possibility of
constructing a transformation group with tensor and spinor generators which
does not at the same time include coordinate transformations.

Let us consider the field equations
\begin{equation}
\left( \gamma _\mu \partial _\mu +\frac 12\left( m_1+m_2\right) \right)
G(x)+\frac 12\left( m_2-m_1\right) \gamma _5G(x)\gamma _5=0  \label{82}
\end{equation}

where $\gamma _\mu $ are the Dirac matrices, and the matrix $G(x)$ is
\begin{equation}
G(x)=\psi _0(x)I_4-\psi _\mu (x)\gamma _\mu +\frac 12\psi _{\mu \nu
}(x)\gamma _{[\mu }\gamma _{\nu ]}+i\widetilde{\psi }_\mu (x)\gamma _\mu
\gamma _5-i\widetilde{\psi }_0(x)\gamma _5  \label{83}
\end{equation}

The quantities $\psi _0(x)$, $\psi _\mu (x)$, $\psi _{\mu \nu
}(x),$ $ \widetilde{\psi }_\mu (x)$, and $\widetilde{\psi }_0(x)$
in Eq. (83) are, respectively, a scalar, a vector, an
antisymmetric tensor, a pseudovector and a pseudoscalar; under the
Lorentz group, $G(x)$ transforms as follows:
\begin{equation}
G(x)\rightarrow G^L(x)=SG(x)S^{-1}\hspace{0.3in}S=\exp \left( \frac 14\omega
_{\mu \nu }\gamma _{[\mu }\gamma _{\nu ]}\right)  \label{84}
\end{equation}

where $\omega _{\mu \nu }$ are the Lorentz group parameters.
Multiplying Eq. (82) successively by the Clifford-algebra elements
$\gamma _A$: $iI_4$, $ \gamma _\mu $, $(1/2)\gamma _{[\mu }\gamma
_{\nu ]}$, $\gamma _\mu \gamma _5$ and $\gamma _5$, and taking the
trace, we find the tensor equations which coincide with Eqs. (9)
including the massive and massless cases. The case of a massless
field corresponds to the choice $m_1=0$, while a massive field (at
$m_1=m_2=m$) is described by an equation of the type
\begin{equation}
\left( \gamma _\mu \partial _\mu +m\right) G(x)=0  \label{85}
\end{equation}

The matrix equation (85) is equivalent to the massive
Dirac-K\"ahler equation. It is obvious that Eq. (85) describes
spinor particles when the $ 4\times 4-$matrix $G(x)$ represents
four bispinors, and that it describes scalar, vector,
antisymmetric tensor, pseudovector and pseudoscalar fields when
$G(x)$ is expanded by (83). For the spinor case, however, $G(x)$
transforms as follows (in this case we use the index $1/2$):
\begin{equation}
G_{1/2}(x)\rightarrow G_{1/2}^L(x)=SG_{1/2}(x)\hspace{0.3in}S=\exp \left(
\frac 14\omega _{\mu \nu }\gamma _{[\mu }\gamma _{\nu ]}\right)  \label{86}
\end{equation}

The Lagrangian corresponding to Eq. (85) is
\begin{equation}
\mathcal{L}=-\frac 12\mbox{tr}\left[ \overline{G}(x)\gamma _\mu
\partial _\mu G(x)-\overline{G}(x)\gamma _\mu
\overleftarrow{\partial _\mu }G(x)+2m \overline{G}(x)G(x)\right]
\label{87}
\end{equation}

where $\overline{G}(x)=\gamma _4G(x)\gamma _4$, the arrow specifies the
direction in which the differential operator acts and tr means the trace of
matrices. After taking the trace in Eq. (87), we arrive at the Lagrangian
which is equivalent to Eq. (63).

Equation (85) is invariant under the following transformations of the matrix
quantity $G(x)$:
\begin{equation}
G(x)\rightarrow G^{\prime }(x)=G(x)D  \label{88}
\end{equation}

The relativistic invariance of Eq. (85) is retained if the matrix $D$
transforms under the Lorentz group as
\begin{equation}
D(x)\rightarrow D^L(x)=SDS^{-1}  \label{89}
\end{equation}

i. e., if the generators of the transformation group (88) are of a tensor
nature with respect to the Lorentz group. In the spinor case, all parameters
of transformation (88) are scalars (not tensors). Therefore the Lorentz
transformations (86) and the transformations of the internal symmetry (88)
commute each other. If we make the Lorentz (86) and symmetry (88)
transformations for the case of spin $1/2$, one gets
\begin{equation}
G_{1/2}(x)\rightarrow G_{1/2}^{L\prime }(x)=SG_{1/2}(x)D  \label{90}
\end{equation}

The commutation of these two groups of transformations is obvious from Eq.
(90). If we put $D=S^{-1}$ in Eq. (90) we arrive at the law of the
transformation of the tensor fields (84). That is why it is possible to
describe the spinor particles with spin $1/2$ by the tensor fields using the
expansion (83). The same conclusion follows from the formalism of Sec. 2.

The requirement that the Lagrangian (87) be invariant under transformations
(88) leads to the condition
\begin{equation}
D\overline{D}=1\hspace{0.3in}\overline{D}(x)=\gamma _4D^{+}\gamma _4
\label{91}
\end{equation}

Writing $D$ in the form
\begin{equation}
D=\exp \left( i\alpha I_4+\beta _\mu \gamma _\mu +\frac 12\Omega _{\mu \nu
}\gamma _{[\mu }\gamma _{\nu ]}+\delta _\mu \gamma _\mu \gamma _5+\rho
\gamma _5\right)  \label{92}
\end{equation}

where $iI_4$, $\gamma _\mu $, $\frac 12\gamma _{[\mu }\gamma _{\nu
]}$, $ \gamma _\mu \gamma _5$ and $\gamma _5$ are the generators
of the group $ GL(4,C)$, we find from condition (91) a restriction
on the parameters: $ \alpha ^{*}=\alpha $, $\beta _m^{*}=\beta
_m$, $\beta _4^{*}=-\beta _4$, $ \Omega _{mn}^{*}=\Omega _{mn}$,
$\Omega _{m4}^{*}=-\Omega _{m4}$, $\delta _m^{*}=\delta _m$,
$\delta _4^{*}=-\delta _4$, $\rho ^{*}=\rho $, in accordance with
a singling out of the $SO(4,2)\otimes U(1)$ subgroup (or locally
isomorphic to $U(2,2)$). The subgroup $U(1)$ corresponds to the
gauge transformations that conserve electric current. So the
symmetry group of equation (85) is $GL(4,C)$ and the corresponding
Lagrangian $-$ $ SO(4,2)\otimes U(1)$ for a case of tensor fields.
In the case of spinor fields, the symmetry group is $U(4)$, in
accordance with conclusions of Sec. 3.

From the invariance of the Lagrangian (87) under transformations (88) [(91)
is taken into account] we find conservation laws for quantities of the type
\begin{equation}
\Theta _{\mu A}=\frac 12\mbox{tr}\left( \gamma _\mu G(x)\gamma
_A\overline{G} (x)-\gamma _\mu G(x)\gamma _4\gamma _A^{+}\gamma
_4\overline{G}(x)\right) \label{93}
\end{equation}

where $\gamma _A=$ $iI_4$, $\gamma _\mu $, $(1/2)\gamma _{[\mu }\gamma _{\nu
]}$, $\gamma _\mu \gamma _5$, $\gamma _5$ and $\gamma _A^{+}$ are the
complex conjugated Clifford-algebra elements. The conservation currents (93)
were found in Sec. 3 in another formalism. From the physical standpoint, the
appearance of this symmetry results from a mass degeneracy of the spin
states of the particle which are mixed by transformations (88). The symmetry
is preserved in a nonlinear generalization of equation (85) (equations of
the Heisenberg type):
\begin{equation}
\left( \gamma _\mu \partial _\mu +m\right) G(x)+lG(x)\overline{G}(x)G(x)=0
\label{94}
\end{equation}

where $l$ is the coupling constant.

The matrix $G(x)$ corresponds to a second-rank bispinor $G_\alpha ^\beta $.
If the bispinor satisfies the Dirac equations with respect to each index
simultaneously, we find a system of Bargmann-Wigner equations [61] which
describe a particle with spin $0$ and system of particles with spin $1$.

By jointly analyzing Eqs. (85) and the Dirac equation
\begin{equation}
\left( \gamma _\mu \partial _\mu +m\right) \Psi (x)=0  \label{95}
\end{equation}

which is invariant under phase transformations of wave function
$\Psi (x)$ [$ \Psi (x)\rightarrow \Psi ^{\prime }(x)=\exp (i\theta
)\Psi (x)$], we can construct a symmetry group which incorporates
these phase transformations and transformation (88) as a subgroup.
The system (85), (95) is invariant under the following
transformations:
\[
G^{\prime }(x)=G(x)D+\Psi (x)\cdot \overline{\zeta }
\]
\begin{equation}
\Psi ^{\prime }(x)=\Psi (x)\lambda +G(x)\xi  \label{96}
\end{equation}

where $\Psi (x)\cdot \overline{\zeta }=\left( \Psi _\alpha
(x)\cdot \overline{\zeta }^\beta \right) $ is the matrix-dyad, and
$\overline{\zeta }$ , $\xi $ are bispinor-parameters, and $\lambda
$ is the complex number-parameter. Transformations (96) form a
group with the following parameter composition law:
\[
D^{\prime \prime }=D^{\prime }D+\xi ^{\prime }\cdot
\overline{\zeta } \hspace{0.3in}\overline{\zeta }^{\prime \prime
}=\lambda ^{\prime }\overline{ \zeta }+\overline{\zeta }^{\prime
}D
\]
\begin{equation}
\lambda ^{\prime \prime }=\lambda ^{\prime }\lambda
+\overline{\zeta } ^{\prime }\xi \hspace{0.3in}\xi ^{\prime \prime
}=\xi ^{\prime }\lambda +D^{\prime }\xi  \label{97}
\end{equation}

Under the Lorentz group, the parameters $\overline{\zeta }$ and
$\xi $ transform as bispinors $\overline{\Psi }$ and $\Psi $,
respectively and are constant quantities, independent of the
space-time coordinates. In order to preserve the relationship
between the spin and the statistics, we must require that the
parameters $\xi $ and $\zeta $ anticommute: $\left\{ \xi _\alpha
,\xi _\beta \right\} =\left\{ \overline{\zeta }^\alpha ,\overline{
\zeta }^\beta \right\} =0$. The need for this condition can be
seen directly from the commutation relations for boson and fermion
fields and from the explicit form of transformations (96).

To establish the group structure of transformations (96) it is convenient to
use a $20-$component column function $\Phi (x)$ whose first components are
formed by the elements of the lines of the matrix $G_\alpha ^\beta (x)$ (in
the order of an alternation of lines and of the elements in them). The other
four components correspond to the wave function $\Psi (x),$ so
\begin{equation}
\Phi (x)=\left(
\begin{array}{c}
G_\alpha ^\beta (x) \\
\Psi _\alpha (x)
\end{array}
\right)  \label{98}
\end{equation}

A direct check confirms the following form for writing transformations (96):
\begin{equation}
\Phi ^{\prime }(x)=\left( I_4\otimes B\right) \Phi (x)\hspace{0.3in}B=\left(
\begin{array}{cc}
D^T & \overline{\zeta }\downarrow \\
\overrightarrow{\xi } & \lambda
\end{array}
\right)  \label{99}
\end{equation}

where $\overline{\zeta }\downarrow $ is a column, and
$\overrightarrow{\xi }$ is a row and $\left( D^T\right) _{\alpha
\beta }=D_{\beta \alpha }$. Under the condition tr$\left(
D^T\right) =\lambda $, the transformations in (99) correspond to
the graduated Lie algebra $SL(4\mid 1)$, in the notation of [62].
The form in (99) for the transformations (96) is of a standard
type [63], where $D^T$ and $\lambda $ are even elements of a
Grassmann algebra, and $\xi $ and $\overline{\zeta }$ are odd
elements (i.e., anticommuting elements).

A fundamental distinction between this symmetry and the ``ordinary''
supersymmetry, with tensor and spinor generators, is that the corresponding
superalgebra is closed without appealing to the generators of a Poincar\'e
group.

By analogy with the description of fields with a maximum spin of $1$, fields
with a maximum spin of $s/2$ can be described by the equations
\begin{equation}
\left( \gamma _\mu \partial _\mu +m\right) G_{\alpha _1\alpha _2...\alpha
_s}=0  \label{100}
\end{equation}

Dirac equations (95) and Eq. (85) constitute a particular case of Eq. (100),
with $(G_{\alpha _1})=\Psi (x)$ and $(G_{\alpha _1\alpha _2})=G(x)$. A
particle with a maximum spin of $3/2$ and a rest mass is described, for
example, by the function $G_{\alpha _1\alpha _2\alpha _3}$. These
internal-symmetry transformations can be generalized immediately to the case
of particles with maximum spins of $1$ and $3/2$, which are the particles of
the greatest physical interest.

What possible physical applications could this symmetry have? In
an analysis of systems consisting of two (mesons) and three
(baryons) quarks in the $s$ state, and in the absence of a
spin-spin interaction, the hadrons may be described as multispin
particles with a rest mass. In this case the symmetry under
consideration holds rigorously, and the hadron interactions can be
associated with the internal-symmetry group $SO(3,1)\otimes
SU(3)$, where $ SO(3,1)$ is the internal symmetry group, which
forms with the Poincar\'e group a semidirect product. As has been
mentioned in the literature [64] the latter property is a
necessary feature of strong-interaction symmetry groups with
incorporated quark spin. From this standpoint the considered
supersymmetry corresponds to an internal symmetry of a field
theory which incorporates, along with composite particles, their
structural components. A further study of this symmetry will be
required, of course, to take into account its possible extension
to interacting fields.

\section{Non-Abelian tensor gauge theory}

We will show the possibility of constructing a gauge model of interacting
Dirac-K\"ahler fields where the gauge group is the noncompact group $SO(4,2)$
under consideration [65].

The starting point in the introduction of Yang-Mills fields is the
localization of parameters of the symmetry group, the transformations of
which do not affect the space-time coordinates. We consider fields $\Psi (x)$
that possess certain transformation properties under the Lorentz group and
that may be transformed under a certain representation of the internal
symmetry group (usually compact and semisimple of type $SU(n)$). For
example, QCD considers the fermionic fields of spin $1/2$, which are
transformed under the fundamental representation of $SU(3)$, in which the
colored quarks are the principal objects. In other words the concept that
there exist some internal quantum numbers (isospin, colour, etc.) is a
requisite physical element of non-Abelian gauge theory. However, it has been
a long-standing problem to describe particles without involving internal
(``isotopic'') spaces (for example, the works of Heisenberg, D\"urr [66, 67]
and Budini [68]), but using instead an adequate generalization of the known
relativistic wave equations (RWE) (see reference [69] and other references
below). It becomes imperative today, for an approach of this kind, to imply
the concept of a non--Abelian gauge field$-$the carrier of interactions. The
possibility of constructing the gauge theory is inherent in the theory of
RWE. The theory might be based on the concept of a multispin or,
equivalently, of a particle having several spin states. The equation remains
coupled$-$i. e., it describes, as the ordinary Dirac equation does, a
particle (and antiparticle) having a set of states, rather than a set of
particles, as is the case with, for example, the equation corresponding to
the direct sum of the Dirac equations. Transformations of internal symmetry
groups result in a mixture of states related to different values of the spin
squared operator. Their localization leads to non-Abelian gauge fields
having multispin $0,1,2$. In this case the dynamics of the interaction of
particles with multispin $0,1$ are associated with the change of their state
through the exchange of particles with maximum spin $2$. The theory
constructed in this manner represents a space-time analogue of gauge theory
with internal symmetry.

An attractive possibility is to describe quarks by the Dirac-K\"ahler
equations. The authors [43-50] used lattice version of the Dirac-K\"ahler
equations describing fermions by inhomogeneous differential forms. This is
equivalent to introducing a set of antisymmetric tensor fields of arbitrary
rank for describing the fermion matter fields. As shown in the Introduction,
we arrive at the Dirac-K\"ahler formulation which includes a scalar, a
vector, an antisymmetric tensor, a pseudovector and a pseudoscalar field.
Here we consider the continuum case of the equations and introduce
non-Abelian tensor gauge fields (gluon fields) for interacting quarks. In
this point of view, quarks possess the multi-spin $0,1$. As we have already
shown, in the continuum case the equation for the $16-$component Dirac
equation can be reduced to four independent $4-$component Dirac equations.
We proceed to use the language of tensor fields to formulate non-Abelian
tensor gauge theory.

The requirement that the Lagrangian (54) be invariant under local
transformations (60) leads to the necessity of introducing a compensating
field $A_\mu ^B$, where the index $B$ is ``internal'' (in our case it
represents a set of tensor indices specifying a scalar, a four-vector, a
skew second-rank tensor, an axial four-vector and a pseudoscalar). The gauge
invariant Lagrangian has the known form:
\begin{equation}
\mathcal{L}=-\overline{\Psi }(x)\left[ \Gamma _\mu \left( \partial _\mu
-ieA_\mu -gA_\mu ^BI^B\right) +m\right] \Psi (x)-\frac 14\mathcal{F}_{\mu
\nu }^2-\frac 14\left( F_{\mu \nu }^B\right) ^2  \label{101}
\end{equation}

where
\[
\mathcal{F}_{\mu \nu }=\partial _\mu A_\nu -\partial _\nu A_\mu
\]
\begin{equation}
F_{\mu \nu }^B=\partial _\mu A_\nu ^B-\partial _\nu A_\mu ^B-\frac
12gc_{CD}^BA_{[\mu }^CA_{\nu ]}^D  \label{102}
\end{equation}

and $c_{CD}^B$ are the structure constants of $SO(4,2)$ group;
$\mathcal{F} _{\mu \nu }$, $F_{\mu \nu }^B$ are the strengths of
the electromagnetic and ``gluon'' fields, respectively; $g$ is the
``gluon'' coupling constant and $ B=\{\mu ,$ $[\alpha \beta ],$
$\widetilde{\mu },$ $\widetilde{0}\}$ . The localization of the
$U(1)$ group produced the electromagnetic field with
four-potential $A_\mu $.

The corresponding wave equations which follow from the Lagrangian (101) are
\begin{equation}
\partial _\nu \mathcal{F}_{\mu \nu }=J_\mu  \label{103}
\end{equation}
\begin{equation}
\partial _\nu F_{\mu \nu }^B+gc_{DC}^BA_\nu ^CF_{\mu \nu }^D=J_\mu ^B
\label{104}
\end{equation}
\begin{equation}
\left[ \Gamma _\mu \left( \partial _\mu -ieA_\mu -gA_\mu ^BI^B\right)
+m\right] \Psi (x)=0  \label{105}
\end{equation}

where $J_\mu =e\overline{\Psi }\Gamma _\mu \Psi $, $J_\mu
^B=g\overline{\Psi }\Gamma _\mu I^B\Psi $; $I^B=\overline{\Gamma
}_\mu $, $(1/4)\overline{ \Gamma }_{_{[\mu }}\overline{\Gamma
}_{\nu ]}$, $\overline{\Gamma }_5$, $ \overline{\Gamma
}_5\overline{\Gamma }_\mu $ (see Eq. (57)). The conservation
current for the non-Abelian fields is
\begin{equation}
\widehat{J}_\mu ^B=J_\mu ^B-gc_{DC}^BA_\nu ^CF_{\mu \nu }^D  \label{106}
\end{equation}

In the general case the gauge field multiplets $A_\nu ^C$ include
a second-rank tensors $A_\nu ^\alpha $, $A_\nu ^{\widetilde{\alpha
}}$, a third-rank tensor antisymmetric over two indices $A_\nu
^{[\mu \nu ]}$, and an axial four-vector $A_\nu ^{\widetilde{0}}$.
The structure constants $ c_{CD}^B$ transform in the tensor
representation of the Lorentz group. The gauge fields carry the
maximal spin $2$. Indeed, the second-rank tensor $ A_\nu ^\alpha $
is transformed on the following superposition of the irreducible
representations of the Lorentz group:
\begin{equation}
\left( \frac 12,\frac 12\right) \otimes \left( \frac 12,\frac 12\right)
=\left( 0,0\right) \oplus \left( 0,1\right) \oplus \left( 1,0\right) \oplus
\left( 1,1\right)  \label{107}
\end{equation}

corresponding to the fields with the spins $0$, $1$, $2$. The third-rank
tensor $A_\nu ^{[\mu \nu ]}$ realizes the representations
\begin{equation}
\left( \frac 12,\frac 12\right) \otimes \left[ \left( 0,1\right) \oplus
\left( 1,0\right) \right] =\left( \frac 12,\frac 12\right) \oplus \left(
\frac 12,\frac 32\right) \oplus \left( \frac 32,\frac 12\right) \oplus
\left( \frac 12,\frac 12\right)  \label{108}
\end{equation}

which also contain fields with spins $0$, $1$, $2$; hence we come to the
same conclusion about the spin of the gauge fields.

We can also consider the localization of some subgroups of the
total group $ SO(4,2).$ Then the Lagrangian (101) will be
invariant under the local transformations of this subgroup. The
main requirement is to extract the subgroup in a relativistic
manner. We suggest the following subgroups and their corresponding
generators
\[
SO(3,1)-\left\{ I_{[\mu \nu ]}\right\} \hspace{0.3in}SO(3,2)-\left\{ I_{[\mu
\nu ]},I_\alpha \right\}
\]
\begin{equation}
SO(4,1)-\left\{ I_{[\mu \nu ]},\widetilde{I}_\alpha \right\}
\hspace{0.3in} GL(1,R)-\left\{ \widetilde{I}\right\}  \label{109}
\end{equation}

The possibility of constructing a dynamic theory with all the main
properties of gauge theories, but based upon notions of space-time rather
than on new internal quantum numbers, is of evident interest.

The absolute group symmetry $G$ corresponds to the semidirect multiplication
of the Poincar\'e group $P$ on the internal symmetry group $D$ ($SO(4,2)$
group):
\begin{equation}
G=P\cdot D  \label{110}
\end{equation}

and the transformations of the symmetry group ($D-$group) commute with the
transformations of the subgroup of four-translations $T_4$. Following [70],
it is possible to define the ``auxiliary'' Poincar\'e group $P^{\prime }$,
which is isomorphic to $P,$ by the relationships:
\begin{equation}
P^{\prime }=\left\{ L_{\mu \nu }^{\prime }\right\} \cdot
T_4\hspace{0.3in} L_{\mu \nu }^{\prime }=L_{\mu \nu }-I_{\mu \nu }
\label{111}
\end{equation}

where $I_{\mu \nu }$ are the generators of the internal symmetry group $%
SO(3,1)$ (see (57), (60)). As a result. we have
\begin{equation}
G=P\cdot D=P^{\prime }\otimes D  \label{112}
\end{equation}

i. e., the absolute group symmetry $G$ represents the direct
product of the auxiliary Poincar\'e group $P^{\prime }$ and the
internal symmetry group $D$ , taking into account that $\left[
L_{\mu \nu }^{\prime },I_{\mu \nu }\right] =\left[ T_4,I_{\mu \nu
}\right] =0$. In our case of the Dirac-K\"ahler equation, $I_{\mu
\nu }=(1/4)(\overline{\Gamma }_\mu \overline{\Gamma }_\nu
-\overline{\Gamma }_\nu \overline{\Gamma }_\mu )$ and
\begin{equation}
L_{\mu \nu }^{\prime }=\frac 14\left( \Gamma _\mu \Gamma _\nu -\Gamma _\nu
\Gamma _\mu \right)  \label{113}
\end{equation}

The generators of the ``auxiliary'' Lorentz group, $L_{\mu \nu
}^{\prime }$, commute with the generators of the internal symmetry
group, $I_{AB}$ (58), and the wave function transforms in the
spinor representation of the group $ \left\{ L_{\mu \nu }^{\prime
}\right\} $. This confirms that the Dirac-K\"ahler equation
describing a set of antisymmetric tensor fields by the
inhomogeneous differential forms, can describe spin $1/2$
particles (pseudoscalar and pseudovector fields are equivalent to
the antisymmetric tensors of fourth and third ranks,
respectively). The non-Abelian gauge theory under consideration is
an analogy to the ordinary non-Abelian gauge theory of spin $1/2$
particles interacting via gluon fields with internal symmetry
group $SO(6)$ (or $SU(4)$). However in our case we have noncompact
gauge group $SO(4,2)$ (or $SU(2,2)$) which requires the
introduction of an indefinite metric.

\section{Conclusion}

We have considered Dirac-K\"ahler equations which can be represented as the
direct sum of four Dirac equations. The main feature of such scheme is the
presence of the additional symmetry associated with non-compact group in the
Minkowski space. The transformations of this group mix fields with different
spins and they do not commute with the Lorentz transformations. As a result,
the group parameters realize tensor representations of the Lorentz group.
This kind of symmetry differs from the colour and flavour symmetries of QCD
and supersymmetry. At the same time it was shown that the Dirac-K\"ahler
fields allow us to introduce graduated groups with tensor and spinor
parameters without including coordinate transformations.

The field scheme considered allow us to construct gauge theories with
different non-compact groups $O(3,1)$, $O(4,2)$, $O(3,3)$, where ``gluon''
fields carry spins $0$, $1$, $2$. Some of these sup-groups become compact
groups in the Euclidean space-time. The theory constructed represents a
space-time analogue of gauge theory with internal symmetry but there is a
difficulty arising from the presence of an indefinite metric. The field
schemes considered can be applied to a construction of quark models or for
the classification of hadrons by non-compact groups (see [72]), and
possibly, for studying sub-quark matter.

The calculated density matrices (matrix-dyads) for fields and the method of
computing the traces of $16-$dimensional Petiau-Duffin-Kemmer matrix
products allow us to make evaluations of different physical quantities in a
covariant manner.

\subsection{8. Appendix}

A method of computing the traces of $16-$dimensional Petiau-Duffin-Kemmer
matrix products will be considered [71].

In order to compare the $16-$component model of vector fields (including
scalar states) with the Proca and Petiau-Duffin-Kemmer theories we consider
here the density matrices in the form of matrix-dyads (40) for pure spin
states. Taking into account Eqs. (16) it is possible to have the simpler
expressions by using equalities
\[
\frac{\sigma ^2}2p^{(+)}=p^{(+)}\frac{\sigma
^2}2=p^{(1)}\hspace{0.3in}\frac{ \sigma
^2}2p^{(-)}=p^{(-)}\frac{\sigma ^2}2=p^{(\widetilde{1})}
\]
\begin{equation}
\left( 1-\frac{\sigma ^2}2\right)
p^{(+)}=p^{(\widetilde{0})}\hspace{0.3in} \left( 1-\frac{\sigma
^2}2\right) p^{(-)}=p^{(0)}  \label{114}
\end{equation}

where $p^{(1)}=p_\mu \beta _\mu ^{(1)},$
$p^{(\widetilde{1})}=p_\mu \beta _\mu ^{(\widetilde{1})},$
$p^{(0)}=p_\mu \beta _\mu ^{(0)}$, $p^{(\widetilde{ 0})}=p_\mu
\beta _\mu ^{(\widetilde{0})}$. The relationships (114) are
obtained by using Eqs. (16), (24) and the expression for the
squared spin operator:
\begin{equation}
\sigma ^2=\left[ \frac i{2m}\epsilon _{\mu \nu \alpha \beta }\frac 14\left(
\Gamma _\mu \Gamma _\nu -\Gamma _\nu \Gamma _\mu +\overline{\Gamma }_\mu
\overline{\Gamma }_\nu -\overline{\Gamma }_\mu \overline{\Gamma }_\nu
\right) \right] ^2  \label{115}
\end{equation}

Taking into account relations (114), the projection matrix-dyads (40) are
transformed into
\begin{equation}
\Delta ^{(1)}=\frac 1{4m^2}ip^{(1)}\left( ip^{(1)}-\varepsilon
m\right) \sigma _p^{(1)}\left( \sigma _p^{(1)}+s_p\right) =\Psi
^{(1)}\cdot \overline{ \Psi }^{(1)}  \label{116}
\end{equation}
\begin{equation}
\Delta ^{(\widetilde{1})}=\frac 1{4m^2}ip^{(\widetilde{1})}\left(
ip^{( \widetilde{1})}-\varepsilon m\right) \sigma
_p^{(\widetilde{1})}\left( \sigma _p^{(\widetilde{1})}+s_p\right)
=\Psi ^{(\widetilde{1})}\cdot \overline{\Psi }^{(\widetilde{1})}
\label{117}
\end{equation}
\begin{equation}
\Delta _0^{(1)}=\frac 1{2m^2}ip^{(1)}\left( ip^{(1)}-\varepsilon m\right)
\left( 1-\sigma _p^{(1)2}\right) =\Psi _0^{(1)}\cdot \overline{\Psi }_0^{(1)}
\label{118}
\end{equation}
\begin{equation}
\Delta _0^{(\widetilde{1})}=\frac
1{2m^2}ip^{(\widetilde{1})}\left( ip^{(
\widetilde{1})}-\varepsilon m\right) \left( 1-\sigma
_p^{(\widetilde{1} )2}\right) =\Psi _0^{(\widetilde{1})}\cdot
\overline{\Psi }_0^{(\widetilde{1} )}  \label{119}
\end{equation}
\[
\Delta ^{(0)}=\frac 1{2m^2}ip^{(0)}\left( ip^{(0)}-\varepsilon m\right)
\left( 1-\sigma _p^{(0)2}\right)
\]
\begin{equation}
=\frac 1{2m^2}ip^{(0)}\left( ip^{(0)}-\varepsilon m\right) =\Psi ^{(0)}\cdot
\overline{\Psi }^{(0)}  \label{120}
\end{equation}
\[
\Delta ^{(\widetilde{0})}=\frac 1{2m^2}ip^{(\widetilde{0})}\left(
ip^{( \widetilde{0})}-\varepsilon m\right) \left( 1-\sigma
_p^{(\widetilde{0} )2}\right)
\]
\begin{equation}
=\frac 1{2m^2}ip^{(\widetilde{0})}\left( ip^{(\widetilde{0})}-\varepsilon
m\right) =\Psi ^{(\widetilde{0})}\cdot \overline{\Psi }^{(\widetilde{0})}
\label{121}
\end{equation}

where we also used here the equalities:
\[
p^{(0)}\left( 1-\sigma _p^{(0)2}\right) =p^{(0)}\hspace{0.3in}p^{(1)}\sigma
_p=p^{(1)}\sigma _p^{(1)}=-\frac i{\mid \mathbf{p}\mid }p^{(1)}\epsilon
_{abc}p_a\beta _b^{(1)}\beta _c^{(1)}
\]

and so on.

The matrix-dyads (116), (118) correspond to the vector state,
(117), (119) - to the pseudovector state, (120) - to the scalar
state, and (121) - to the pseudoscalar state. Eventually Eqs.
(116)-(121) coincide with solutions of $ 10-$component (for vector
and pseudovector states) and $5-$component (for scalar and
pseudoscalar states) free Petiau-Duffin-Kemmer equations.

It is known that cross-sections for the scattering processes are summed to
evaluate the transition probabilities for a particle going from the initial
to the final states. These probabilities are proportional to the squared
module's of the matrix elements which can be written as (see for example
[49, 50]):
\begin{equation}
\mid M\mid ^2=e^2\mbox{tr}\left\{ Q\Pi _1\overline{Q}\Pi
_2\right\} \label{122}
\end{equation}

where $Q$ is the vertex operator, $\overline{Q}=\eta Q^{+}\eta $
($\eta =\Gamma _4\overline{\Gamma }_4$ is the Hermitianizing
matrix, $Q^{+}$ is the Hermite conjugated operator), the
matrix-dyads $\Pi _1$, $\Pi _2$ correspond to initial and final
states, respectively. Therefore we need the traces of $
16-$dimensional Petiau-Duffin-Kemmer matrix products to calculate
some processes with the presence of vector and scalar fields.

It is easy to verify that the property of traces of the $16\times 16-$
Petiau-Duffin-Kemmer matrices $\beta _\mu ^{(\pm )}$ (16)
\[
\mbox{tr}\left\{ \beta _{\mu _1}\beta _{\mu _2}...\beta _{\mu
_n}\right\} = \mbox{tr}\left\{ \left( P\beta _{\mu _n}\beta _{\mu
_1}P\right) \left( P\beta _{\mu _2}\beta _{\mu _3}P\right)
...\left( P\beta _{\mu _{n-2}}\beta _{\mu _{n-1}}P\right) \right\}
\]
\begin{equation}
+\mbox{tr}\left\{ \left( P\beta _{\mu _1}\beta _{\mu _2}P\right)
\left( P\beta _{\mu _3}\beta _{\mu _4}P\right) ...\left( P\beta
_{\mu _{n-1}}\beta _{\mu _n}P\right) \right\}  \label{123}
\end{equation}

is valid, where $P=\varepsilon ^{0,0}+\varepsilon
^{\widetilde{0},\widetilde{ 0}}+(1/2)\varepsilon ^{[\mu \nu ],[\mu
\nu ]}$ is the projection operator. The analogous identity was
derived in [45] for the $5\times 5-$ and $ 10\times
10-$Petiau-Duffin-Kemmer matrices. The trace of the odd-numbered
matrices $\beta _\mu ^{(\pm )}$ is equal to zero. Eq. (123) is
valid for any Petiau-Duffin-Kemmer matrix given by (16). Using the
properties of the entire matrix algebra $\varepsilon ^{A,B}$ we
find
\[
P\beta _\mu ^{(-)}\beta _\nu ^{(+)}P=\varepsilon ^{0,[\mu \nu ]}+\frac
12e_{\nu \mu \rho \omega }\varepsilon ^{[\rho \omega ],\widetilde{0}}
\]
\[
P\beta _\mu ^{(+)}\beta _\nu ^{(-)}P=\varepsilon ^{[\mu \nu ],0}+\frac
12e_{\mu \nu \rho \omega }\varepsilon ^{\widetilde{0},[\rho \omega ]}
\]
\[
P\beta _\mu ^{(+)}\beta _\nu ^{(+)}P=\delta _{\mu \nu }\varepsilon
^{ \widetilde{0},\widetilde{0}}+\varepsilon ^{[\rho \mu ],[\rho
\nu ]}
\]
\[
P\beta _\mu ^{(-)}\beta _\nu ^{(-)}P=\delta _{\mu \nu }\varepsilon
^{0,0}+\frac 14e_{\lambda \mu \rho \omega }e_{\lambda \nu \sigma \alpha
}\varepsilon ^{[\rho \omega ][\sigma \alpha ]}
\]
\[
P\beta _\mu ^{(0)}\beta _\nu ^{(+)}P=P\beta _\mu ^{(-)}\beta _\nu
^{(1)}P=\varepsilon ^{0,[\mu \nu ]}
\]
\[
P\beta _\mu ^{(+)}\beta _\nu ^{(0)}P=P\beta _\mu ^{(1)}\beta _\nu
^{(-)}P=\varepsilon ^{[\nu \mu ],0}
\]
\begin{equation}
P\beta _\mu ^{(0)}\beta _\nu ^{(-)}P=P\beta _\mu ^{(-)}\beta _\nu
^{(0)}P=\delta _{\mu \nu }\varepsilon ^{0,0}  \label{124}
\end{equation}

The relations (123), (124) with the help of the equalities tr$\left\{
\varepsilon ^{A,B}\right\} =\delta _{A,B}$, $\delta _{[\mu \nu ][\rho \sigma
]}=\delta _{\mu \rho }\delta _{\nu \sigma }-\delta _{\mu \sigma }\delta
_{\nu \rho }$, $\varepsilon ^{A,B}\varepsilon ^{C,D}=\delta _{BC}\varepsilon
^{A,D}$, $\left( \varepsilon ^{A,B}\right) _{CD}=\delta _{AC}\delta _{BD}$
allow us to find traces of any Petiau-Duffin-Kemmer matrices.

\end{document}